\documentclass[journal]{IEEEtran}

\usepackage{graphicx}
\usepackage{amsmath,amssymb}
\usepackage{booktabs}
\usepackage{tabularx}
\usepackage{cite}
\usepackage{siunitx}

\usepackage{xcolor}

\usepackage{booktabs,tabularx,array}
\newcolumntype{L}[1]{>{\raggedright\arraybackslash}p{#1}}
\newcolumntype{C}[1]{>{\centering\arraybackslash}p{#1}}
\newcolumntype{Y}{>{\raggedright\arraybackslash}X}

\title{Advanced Superdirective Antennas}

\author{Alex~Krasnok, \textit{IEEE Senior Member}
\thanks{Alex Krasnok is with the Department of Electrical and Computer Engineering, Florida International University, Miami, FL 33174 USA.}%
\thanks{The author acknowledges financial support from the U.S. Department of Energy (DoE) and the U.S. Air Force Office of Scientific Research (AFOSR).}%
}

\begin{document}
\maketitle

\begin{abstract}
Superdirective (supergain) antennas aim to produce a narrow main beam from radiators that are electrically small compared with the wavelength. Instead of enlarging the physical aperture, they rely on strongly coupled currents, near-field energy storage, and controlled modal interference so that a compact structure radiates with enhanced directivity. This review emphasizes link-relevant evaluation and reporting: realized gain referenced to a stated impedance plane, clearly stated bandwidth definitions (impedance and performance), and robustness to fabrication spread and platform/environmental loading. Two practical implementation routes are surveyed. The first uses resonant, tightly coupled arrays, including fully driven arrays and single-chain designs based on parasitic or reactively loaded elements. The second uses single-body radiators that enforce a targeted mixture of multipoles or resonant/characteristic modes with one or a few feeds, including symmetry-broken dielectric resonators and mixed electric--magnetic designs. Across RF, microwave, and optical regimes, the same penalties recur as superdirectivity is pushed: reduced radiation resistance, rapid impedance variation, narrow usable bandwidth, and strong sensitivity to small perturbations. Beyond geometric synthesis and multi-resonant stacking, the review highlights emerging levers that can shift these trade-offs in specific system contexts: low-loss materials and cryogenic operation to improve efficiency and frequency stability, and time-varying loading and matching (Floquet/parametric approaches) that can relax linear time-invariant bandwidth constraints, at the cost of added control complexity and spectral conversion.
\end{abstract}


\section{Introduction}
High directivity is usually obtained by increasing electrical aperture. Reflectors, lenses, leaky-wave antennas, and large arrays narrow the beam by distributing radiating currents over several wavelengths \cite{Balanis2016}. In contrast, electrically small antennas, commonly characterized by the size factor $ka$ with $k=2\pi/\lambda$ and $a$ the radius of the smallest enclosing sphere, tend to radiate dipole-like patterns and face strong constraints on matching and bandwidth \cite{Wheeler1947,Chu1948,YaghjianBest2005}. {Superdirective antennas pursue the opposite design choice: they aim to exceed the directivity expected at a given $ka$ by shaping currents and near fields, not by expanding the aperture. Here, we use \emph{superdirectivity} operationally to mean that the peak directivity $D_{\max}$ exceeds a practical ``normal-directivity'' reference for the same $ka$ (e.g., Harrington's estimate $D_H(ka)\approx (ka)^2+2ka$), and we use \emph{supergain} when referring specifically to the classical end-fire array literature.}

The possibility of supergain is not new. Classic end-fire array analyses established that, in idealized settings, directivity can increase rapidly as inter-element spacing shrinks and excitation becomes strongly nonuniform \cite{HansenWoodyard1938,Uzkov1946}. The historical barrier has been practical rather than conceptual: as superdirectivity is pushed harder, radiation resistance decreases, stored energy grows, and the required excitation becomes increasingly sensitive to small errors, so loss, mismatch, and tolerances can erase the expected gain advantage in realizable systems \cite{Chu1948,YaghjianBest2005,Harrington1960}. {This tension motivates two themes that recur throughout the literature: first, fair comparison requires link-relevant metrics (realized gain and total efficiency at a stated bandwidth and size), not peak directivity alone; second, ``working'' superdirectivity is typically narrowband unless additional degrees of freedom are introduced to broaden operation or to tune the operating point.}

\begin{figure*}[t]
\centering
\fbox{\begin{minipage}{0.965\textwidth}
\small
\textbf{Box 1. Basic concepts.}
Directivity $D(\theta,\phi)$ compares radiation intensity in a given direction to the average radiation intensity over all directions. {Gain $G(\theta,\phi)$ is defined like directivity but referenced to the \emph{net power accepted} by the antenna at its reference plane, $P_{\mathrm{acc}}$; it therefore excludes reflection (impedance-mismatch) loss. Let $P_{\mathrm{inc}}$ denote the incident power at the antenna reference plane (for a stated reference impedance) and $\Gamma$ the corresponding reflection coefficient, so $P_{\mathrm{acc}}=P_{\mathrm{inc}}(1-|\Gamma|^2)$.} Polarization mismatch is a link-level factor between two antennas and is not included in $G$. {Realized gain $G_{\mathrm{real}}(\theta,\phi)$ is the gain referenced to $P_{\mathrm{inc}}$ (i.e., reduced by the mismatch factor), so it depends on the reference impedance and the system to which the antenna is connected \cite{IEEE1452013}.}

We use the following identities. The radiation efficiency $\eta_{\mathrm{rad}}$ captures dissipative losses (conductors, dielectrics, lumped components). {With reflection coefficient $\Gamma$ at the antenna reference plane, the mismatch factor is $(1-|\Gamma|^2)=P_{\mathrm{acc}}/P_{\mathrm{inc}}$.} Then
\begin{equation}
G = \eta_{\mathrm{rad}}\,D,\qquad
G_{\mathrm{real}} = (1-|\Gamma|^2)\,G \equiv \eta_{\mathrm{tot}}\,D,
\end{equation}
where $\eta_{\mathrm{tot}}=\eta_{\mathrm{rad}}(1-|\Gamma|^2)$ is the total efficiency used when comparing link performance.

For receiving (polarization matched and under the usual conjugate-match assumptions), the effective area $A_e$ relates to gain as
\begin{equation}
A_e = \frac{\lambda^2}{4\pi}\,G.
\end{equation}
If the receiving system is not impedance matched, an additional mismatch factor must be applied at the receiver reference plane.

Size is reported using $ka=2\pi a/\lambda$ with $a$ taken as the smallest enclosing sphere of the radiating system being compared. {In this review, ``superdirectivity'' is used operationally as $D_{\max}>D_H(ka)$ for the same reported $ka$, with $D_H(ka)\approx (ka)^2+2ka$ used as a practical reference baseline (not a universal bound).} When a ground plane or host platform is essential to operation, the comparison must state whether the platform is included in $a$ (radiator-only size versus system size). Bandwidth is reported in two ways when data permit: an impedance bandwidth (e.g., $|S_{11}|<-10$~dB) and a performance bandwidth (e.g., realized gain within 1~dB of its peak).
\end{minipage}}
\end{figure*}

\begin{figure*}[h]
\centering
\includegraphics[width=0.9\textwidth]{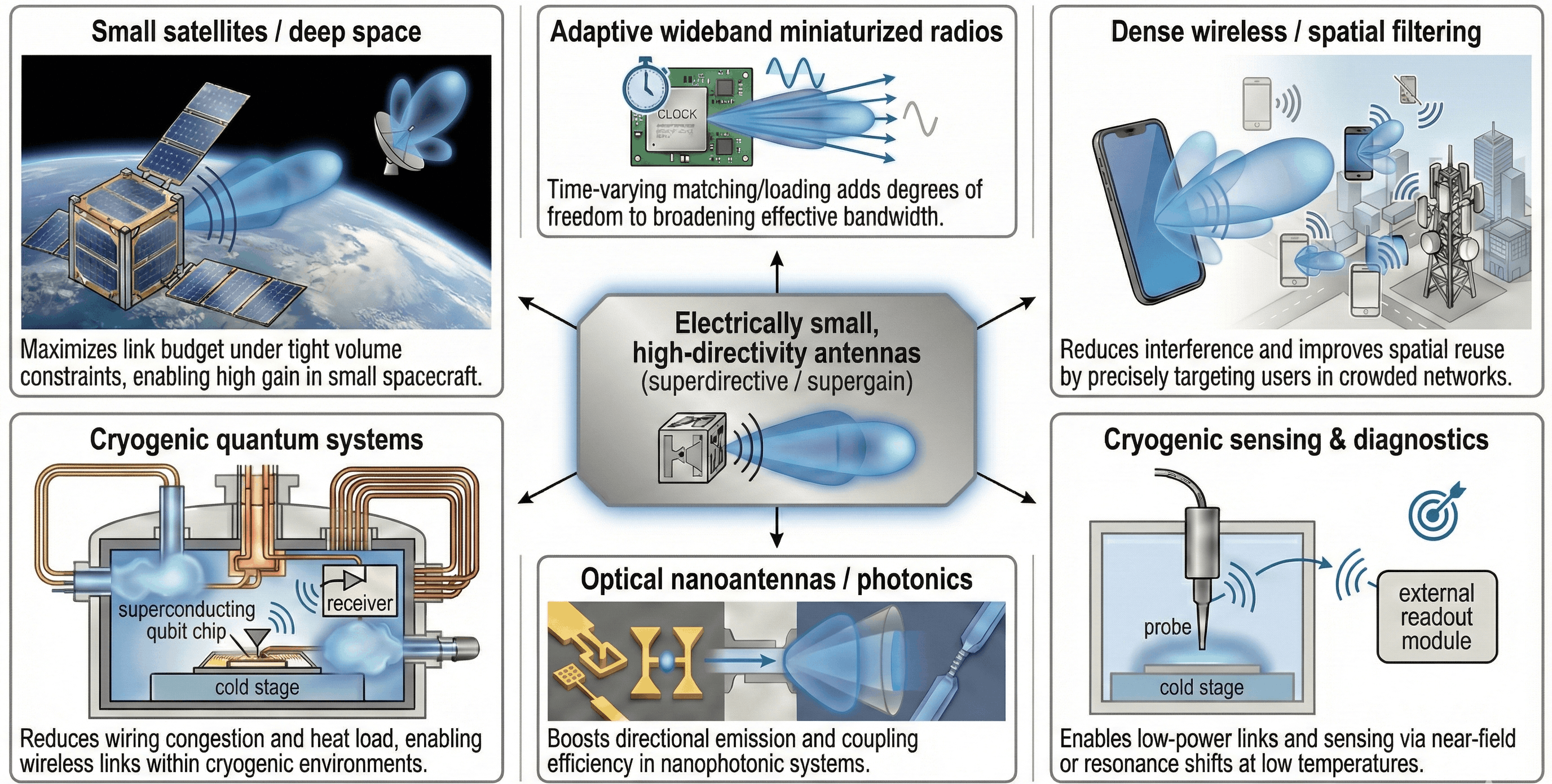} 
\caption{\textbf{Applications of electrically small, high-directivity (superdirective/supergain) antennas.}
Representative system settings where compact directivity can provide link-level value: small satellites/deep space, dense wireless spatial filtering, cryogenic quantum systems, cryogenic sensing/diagnostics, and optical nanoantennas/photonics, along with emerging time-varying matching/loading approaches that can broaden effective operation of electrically small radiators.}
\label{fig:Fig0}
\end{figure*}

Interest has strengthened because the use-cases are concrete and span frequency regimes, as summarized in Fig.~\ref{fig:Fig0}. In small satellites and deep-space links, compact high-gain apertures are valuable when volume and mass are tightly constrained \cite{Abulgasem2021CubesatAntennaReview}. In dense wireless networks, directionality can reduce interference and improve spatial reuse, so even modest directivity from a small form factor can translate into system-level gains \cite{Dai2013DirectionalAntennas}. At optical frequencies, nanoantennas use directional emission and scattering to couple efficiently into finite numerical apertures and guided modes \cite{Novotny2011,Biagioni2012,Kuznetsov2016Science}. Cryogenic platforms add a distinct driver: wireless links inside or across cryostats can reduce wiring congestion and heat load, and low-loss, thermally stable dielectric antennas have been demonstrated down to the 10~K regime \cite{Centritto2025CryoCMOS,TorresKrasnok2026Cryogenics}. Finally, time-varying loading and matching provide an emerging route to broaden effective operation of electrically small antennas beyond linear time-invariant constraints, which is especially relevant when superdirectivity is otherwise confined to narrow resonant peaks \cite{LiMekawyAlu2019PRL,HayranMonticone2023APM}.

\begin{figure*}[t!]
\centering
\includegraphics[width=0.9\textwidth]{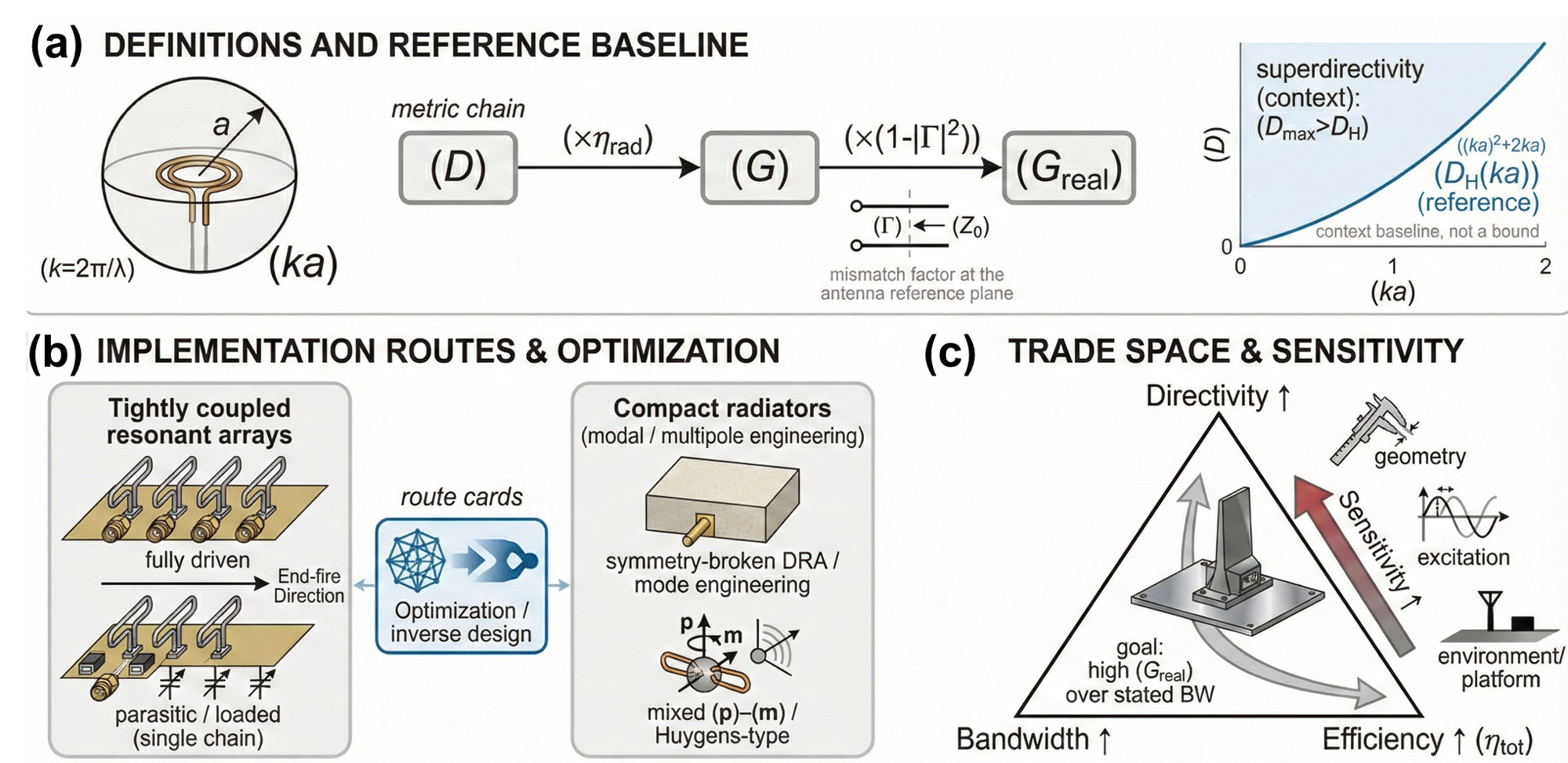}
\caption{\textbf{Reporting baselines for superdirective antennas.}
(a) Definitions and reference baselines used in this review: size factor $ka$ (with $a$ the smallest enclosing-sphere radius), directivity, gain/realized gain, and {a practical ``normal-directivity'' reference $D_H(ka)$ for context.}
(b) Two main implementation routes: tightly coupled resonant arrays (driven or parasitic/loaded) and compact radiators that enforce a desired multipole or modal mixture with few feeds (including dielectric resonator and mixed-multipole designs), with optimization/inverse design as a practical synthesis tool.
(c) The dominant trade space among directivity, bandwidth, and efficiency, and the associated sensitivity to perturbations (geometry, excitation, and environment).}
\label{fig:taxonomy}
\end{figure*}

A practical difficulty is that reported results are often hard to compare across papers and communities. Size may be reported as $ka$ for a radiator alone, or as a footprint including a ground plane or host platform; bandwidth may be reported as an impedance criterion (e.g., $|S_{11}|<-10$~dB) or as a performance criterion (e.g., realized gain within 1~dB of peak); and gain may be reported as directivity, gain, or realized gain without a consistent accounting of mismatch and loss. This review is written to reduce that ambiguity. Box~1 states the definitions used, Fig.~\ref{fig:taxonomy} summarizes the implementation routes, and Table~\ref{tab:sota} collects representative devices together with the quantities needed for fair comparison. {Throughout, we denote the ``normal-directivity'' reference as $D_H(ka)$ to keep notation consistent with directivity.} Throughout the text, representative device sizes are given in wavelength-scaled form (e.g., $ka$ or a $\lambda$-scaled bounding size).

Recent measurements illustrate both what is feasible and what remains costly. Resonant two-element end-fire arrays can achieve measured supergain at separations on the order of $0.1\lambda$--$0.2\lambda$ with electrical size $ka$ on the order of unity, but only over a narrow band and with strong sensitivity to loss and mismatch \cite{Yaghjian2008RadioScience}. Loaded and parasitic multi-element arrays can reach higher directivity at comparable electrical size, yet typically exhibit MHz-scale impedance bandwidths at sub-GHz frequencies and strong dependence on the surrounding platform currents \cite{Haskou2017CRPhys}. Single-feed mode-engineered radiators, such as symmetry-broken dielectric resonators, demonstrate that a compact body can access a higher-order multipole mixture with one feed and produce a narrow, well-defined directive beam \cite{KrasnokAPL2014}. More recently, optimization-driven multi-resonant geometries have shown that the usable band can be widened by stacking several nearby resonances within a subwavelength volume, yielding high measured realized gain over double-digit fractional bandwidth in favorable implementations \cite{Vovchuk2024}. These examples motivate the central message of the paper: superdirectivity is best evaluated as a link-level design choice, using realized gain at a stated bandwidth and size, together with robustness to perturbations.

\section{Physical picture and limits}
Directivity in the direction of maximum radiation is
\begin{equation}
D \;=\; \frac{4\pi\,U_{\max}}{P_{\mathrm{rad}}},
\label{eq:Ddef}
\end{equation}
where $U_{\max}$ is the maximum radiation intensity and $P_{\mathrm{rad}}$ is total radiated power \cite{IEEE1452013,Balanis2016}. For link-relevant comparisons, directivity alone is insufficient because loss and mismatch can dominate. {To avoid ambiguity, we use the IEEE terminology summarized in Box~1: let $P_{\mathrm{inc}}$ be the incident power at the antenna reference plane (for a stated reference impedance) and $P_{\mathrm{acc}}$ the net accepted power, with $P_{\mathrm{acc}}=P_{\mathrm{inc}}(1-|\Gamma|^2)$ where $\Gamma$ is the reflection coefficient. Gain is referenced to $P_{\mathrm{acc}}$ and therefore excludes reflection (mismatch) loss, while realized gain is referenced to $P_{\mathrm{inc}}$ and therefore includes the mismatch factor \cite{IEEE1452013}.} Using radiation efficiency $\eta_{\mathrm{rad}}$ {(defined as $P_{\mathrm{rad}}/P_{\mathrm{acc}}$)} and reflection coefficient $\Gamma$ at the antenna reference plane,
\begin{equation}
G = \eta_{\mathrm{rad}}\,D,\qquad
G_{\mathrm{real}} = (1-|\Gamma|^2)\,G,
\label{eq:gain_def}
\end{equation}
so a large $D$ can translate into modest realized gain if matching and dissipation are poor. {This distinction is especially important in superdirective designs because the same current cancellations that sharpen the pattern can drive the radiation resistance down and make both $\eta_{\mathrm{rad}}$ and $(1-|\Gamma|^2)$ first-order determinants of link benefit.}

A related receiving metric is the effective area. For a polarization-matched receiving antenna under standard matching assumptions,
\begin{equation}
A_{\mathrm{e}} = \frac{\lambda^2}{4\pi}\,G,
\label{eq:Ae}
\end{equation}
which highlights the design goal: realize a large effective area without a physically large aperture \cite{IEEE1452013}. Because $A_e$ is tied to gain, not directivity, loss and matching remain central even when the radiation pattern is sharp. {When mismatch at the antenna terminals is included, it is often convenient to use a realized effective area $A_{\mathrm{e,real}}=(\lambda^2/4\pi)G_{\mathrm{real}}$, or equivalently to apply the factor $(1-|\Gamma|^2)$ at the appropriate reference plane in the receive chain.}

The classical array result is Uzkov’s limit for end-fire directivity of $N$ isotropic radiators as spacing tends to zero, which scales as $N^2$ \cite{Uzkov1946}. The cost is that the required excitations become increasingly large and sensitive to perturbations, and the radiation resistance seen by each port can become very small. This behavior was recognized in early end-fire studies and has been quantified in later analyses and measurements that explicitly track tolerance, mismatch, and efficiency as spacing is reduced \cite{HansenWoodyard1938,Harrington1960,Yaghjian2008RadioScience,Haskou2017CRPhys}. {In practical realizations, the $N^2$ trend is therefore better interpreted as an indicator of an increasingly ill-conditioned excitation problem: small amplitude/phase errors, finite loss, or modest environmental loading can produce large deviations in the achieved pattern and $G_{\mathrm{real}}$.} A useful unifying view is that both array supergain and single-body superdirectivity require strong higher-order spatial content: in arrays this appears as rapidly varying current distributions across closely spaced elements; in single-body designs it appears as controlled excitation of higher multipoles or higher-order resonant modes. {In both cases, the same higher-order content is accompanied by strong reactive near fields and increased stored energy, which provides a physical link between superdirectivity, narrow bandwidth, and sensitivity to nearby objects and platforms.} Figure~\ref{fig:taxonomy}(b) frames these two routes.

For benchmarking, many papers compare against Harrington’s “normal directivity” estimate for a radiator confined to radius $a$,
\begin{equation}
D_{\mathrm{H}}(ka) \approx (ka)^2 + 2ka,
\label{eq:Harrington}
\end{equation}
and treat superdirectivity operationally as exceeding this benchmark for the same $ka$ \cite{Harrington1958,Haskou2017CRPhys}. {We denote this reference consistently as $D_{\mathrm{H}}(ka)$ in the text and figures (rather than $G_{\mathrm{H}}$), since its role here is a directivity baseline for pattern sharpening at fixed electrical size.} This benchmark provides useful context, but it is not a universal bound. Modern convex-optimization results show that achievable gain/directivity depends on the imposed constraints, including loss, self-resonance, and what current distributions are physically controllable for a given structure \cite{GustafssonNordebo2013,GustafssonCapek2019}. For this reason, the paper uses $D_{\mathrm{H}}(ka)$ as a reference point rather than a hard limit and emphasizes reported realized performance whenever available.

Bandwidth penalties are naturally expressed through stored energy and $Q$. In the electrically small regime, Chu’s canonical bound provides a baseline scaling trend for minimum $Q$ with size \cite{Chu1948}, and more general formulations show that $Q$ increases when higher directivity constraints are imposed \cite{YaghjianBest2005,Jonsson2017QDirectivity}. Since small matched bandwidth is linked to large $Q$, narrow impedance bandwidth is expected when a design operates deep in the superdirective regime. {Moreover, because the directive pattern relies on interference among a small set of strongly coupled currents or modes, the \emph{pattern} and $G_{\mathrm{real}}$ can change rapidly with frequency even when $|S_{11}|$ appears acceptable; this is why performance bandwidth (e.g., $G_{\mathrm{real}}$ within 1~dB of peak) can be markedly narrower than an impedance bandwidth criterion in resonant superdirective designs.} Bandwidth extension therefore typically requires added degrees of freedom, such as multiple resonances, additional ports, or accepting reduced peak directivity, consistent with the trade sketched in Fig.~\ref{fig:taxonomy}(c) and summarized across reported devices in Table~\ref{tab:sota}. {Material and miniaturization choices interact with this trade: for example, high-permittivity dielectrics can reduce physical size at fixed $ka$, but often increase field confinement and stored energy, tending to raise $Q$ unless multiple modes are deliberately co-designed to broaden operation.}

Matching constraints are often the immediate practical limiter. Even when a superdirective current distribution exists at one frequency, the input impedance can vary rapidly with frequency, and broadband matching is fundamentally limited for reactive loads by Fano-type results \cite{Fano1950,YaghjianBest2005}. {Because matching networks and tuning elements add loss and can perturb the modal balance that produces superdirectivity, the most informative reporting practice is to state both (i) an impedance bandwidth criterion at the antenna reference plane and (ii) a performance bandwidth criterion based on $G_{\mathrm{real}}$ (or $\eta_{\mathrm{tot}}D$) over the same band.} For this reason, the paper distinguishes impedance bandwidth (e.g., $|S_{11}|<-10$~dB) from performance bandwidth (e.g., realized gain within 1~dB of peak) and reports both when possible.

\section{Engineering routes to superdirectivity}
As summarized in Fig.~\ref{fig:taxonomy}(b), practical superdirective antennas follow two broad engineering routes. The first route uses tightly coupled resonant arrays whose element currents are set by carefully chosen amplitudes and phases, either with multiple feeds or with a single feed plus passive loading. The second route uses compact radiators that force a desired mixture of radiation modes (multipoles or characteristic modes) with one or a few feeds. {Both routes can be viewed as controllability problems: arrays attempt to realize a target element-current vector within a coupled impedance network, while single-body designs attempt to realize a target modal (multipole) coefficient vector within a compact resonator.} Both routes aim to create spatial field variations that are unusually strong for a small radiator, which is why the same designs often become narrowband and sensitive, as sketched in Fig.~\ref{fig:taxonomy}(c). {Inverse design and optimization are increasingly used as synthesis tools for both routes because they can include loss, matching, and bandwidth objectives directly, rather than maximizing $D$ at a single frequency.} Table~\ref{tab:sota} collects representative implementations and the quantities needed for system-level comparison.

The classic array starting point is end-fire synthesis. The Hansen--Woodyard condition \cite{HansenWoodyard1938} sets conventional end-fire phasing, and supergain is obtained by pushing the excitation toward strongly nonuniform, nearly out-of-phase adjacent elements as spacing shrinks. The practical difficulty is that shrinking spacing drives radiation resistance downward and stored reactive energy upward, so realized gain becomes dominated by mismatch and loss. A widely cited experimental benchmark that makes these penalties explicit is the resonant electrically small end-fire array work of Yaghjian \emph{et al.} \cite{Yaghjian2008RadioScience}, which demonstrates that resonance can make supergain measurable in hardware while preserving the expected narrowband and loss sensitivity. {In this regime, reducing dissipative loss is as important as refining excitation, since small series resistances and dielectric losses can consume a large fraction of $P_{\mathrm{acc}}$ when radiation resistance is low; this motivates careful materials and fabrication choices, and in some application niches even cryogenic operation where dielectric loss and stability can be characterized and improved \cite{TorresKrasnok2026Cryogenics}.}

A closely related strategy is to shape the element geometry so that the input resistance is closer to standard feed impedances, reducing the need for extreme external matching. Best \emph{et al.} demonstrate this idea with an end-fire array based on folded monopole elements engineered for higher radiation resistance \cite{Best2008}. The array physics is unchanged, but the system-level penalty can be reduced because less power is dissipated or reflected in the feed and matching network. This theme recurs across the literature: designs that appear similar in directivity can differ substantially in realized gain once matching and loss are accounted for.

Parasitic loading reduces feed count by trading electronics for coupling and load synthesis. A common workflow is to compute the optimal current vector for a fully driven array and then use the array impedance matrix to synthesize load impedances that reproduce those currents when only one (or a subset) of elements is driven \cite{Haskou2015}. This approach can approach driven-array directivity with a single RF chain, but it exposes two practical limits emphasized in \cite{Haskou2017CRPhys}: sensitivity increases rapidly as spacing shrinks, and exact current synthesis can require non-passive (negative-resistance) loads. {Even when passive synthesis is feasible, the equivalent series resistance and bias dependence of reactive loads enter $\eta_{\mathrm{rad}}$ directly, so load loss can dominate $G_{\mathrm{real}}$ in the most aggressive superdirective settings.} Platform coupling becomes unavoidable in many electrically small realizations, because the host ground and nearby conductors reshape the return currents that participate in radiation. PCB-integration studies show that controlling where currents flow on the platform can determine whether the intended superdirective mode is preserved or suppressed \cite{Haskou2016LAWP}. In practice, this means that many “small” superdirective arrays must be co-designed with their environment, and size should be reported as a system property when the platform is essential.

Mode-based design tools are useful when the objective is not only peak directivity but also controlled bandwidth and reduced sensitivity. The theory of characteristic modes \cite{HarringtonMautz1971} and its use in antenna shape synthesis \cite{GarbaczPozar1982} provide a systematic way to identify radiating current modes and their coupling to feeds. {From the reporting perspective, this modal view is also helpful because it clarifies why performance bandwidth can be narrow: the directive pattern is tied to a particular superposition of a few modes whose relative weights can change rapidly with frequency and loading.} In multiport arrays, related modal/network views help connect excitation choices to the radiating subspace and can guide where loading or additional resonances are most effective. Representative designs that aim to broaden operation use wideband unit elements and then optimize the array excitation over a target band, rather than relying on a single narrow supergain peak \cite{Jaafar2018URSI}. The recurring lesson is that claims of “wideband superdirectivity” usually reflect a two-step design: build bandwidth and matching into the element and network first, then use coupling to shape the pattern.

Strongly coupled resonator dimers provide a compact alternative to conventional wire or patch arrays because coupling itself sets the relative phase and amplitude between radiating moments. In split-ring resonator (SRR) dimers, the radiating contributions include both electric and magnetic moments, and the dimer orientation controls how these moments combine in the far field. Vallecchi \emph{et al.} demonstrate a singly driven SRR dimer that approaches the two-dipole superdirective limit, with an overall subwavelength scale and a narrowband directive response set by the coupled resonance \cite{Vallecchi2020}. This class connects naturally to coupled-mode and circuit models and to magnetoinductive-wave concepts in resonant meta-atom chains \cite{Shamonina2002JAP}. Related work replaces SRRs with other magnetically coupled resonators, such as subwavelength helical elements, to obtain compact end-fire superdirectivity with simplified feeding \cite{Petrov2020}. These dimer-based designs illustrate a useful middle ground: they avoid multiport excitation, but they retain the fundamental sensitivity of narrowband coupled resonances.

Single-body superdirectivity avoids multi-element feeding by shaping the internal modal content of one compact radiator. A clear microwave example is the notched dielectric resonator antenna of \cite{KrasnokAPL2014}, where symmetry breaking enables selective excitation of higher-order multipoles that are weak in a homogeneous body, producing a narrow but well-defined directive beam with high radiation efficiency near the operating point. The same geometric principle carries to optical nanoantennas, where a high-index particle with broken symmetry can support a controlled mixture of electric and magnetic multipoles that sets the scattering directionality \cite{KrasnokNanoscale2014}. {These designs emphasize that ``few feeds'' does not imply ``few degrees of freedom'': one feed can excite multiple resonant modes when symmetry is intentionally broken, but the achieved superposition must remain stable under dispersion, fabrication spread, and environmental loading to preserve $G_{\mathrm{real}}$ over the intended band.}

Finally, recent “advanced” directions treat bandwidth and robustness as explicit design objectives. Multi-resonant synthesis and inverse design can trade geometric complexity for usable bandwidth and realized gain by placing several controlled resonances close enough in frequency to avoid a single narrow peak \cite{Vovchuk2024}. This approach does not remove the trade-off in Fig.~\ref{fig:taxonomy}(c); it changes how the trade is managed by increasing the number of degrees of freedom and shifting the central question from “maximize directivity at one frequency” to “maximize realized gain over a specified band under practical constraints.” {This shift also changes what should be reported: peak $D$ alone is rarely predictive of link benefit unless accompanied by $\eta_{\mathrm{rad}}$, $(1-|\Gamma|^2)$, and a performance bandwidth definition that reflects how quickly the optimized modal balance degrades away from the design frequency.} As the design space becomes higher-dimensional, careful reporting becomes more important, because the same optimized structure can look impressive in peak directivity while offering limited link improvement if mismatch, loss, or sensitivity dominate. Table~\ref{tab:sota} is organized to keep these comparisons explicit.

%%%%
\section{State of the art across RF, microwave, and optical regimes}
Figure~\ref{fig:taxonomy}(b) groups practical superdirective antennas into two families: strongly coupled arrays (fully driven or reactively loaded/parasitic) and compact radiators that enforce a targeted mixture of radiation modes with one or a few feeds. Table~\ref{tab:sota} provides a selective snapshot across these routes and across frequency regimes. {The goal is not completeness, but a comparison that is defensible at the link level: whenever possible we track realized gain in dBi, a stated bandwidth definition, and an efficiency quantity (radiation or total).} When a source reports only one bandwidth definition, we label it explicitly as an impedance bandwidth (e.g., $|S_{11}|<-10$~dB) or a performance bandwidth (e.g., realized gain within 1~dB of its peak), because these can differ substantially for resonant superdirective designs.

Two-element resonant end-fire arrays remain the cleanest experimental benchmark because the physics is transparent and the system penalties appear immediately. Yaghjian \emph{et al.} demonstrate electrically small two-element supergain arrays with element spacing on the order of $0.15\lambda$ and overall size $ka$ on the order of unity, achieving measured end-fire gains of roughly {6--7~dBi} over a narrow operating band \cite{Yaghjian2008RadioScience}. Best \emph{et al.} emphasize a complementary engineering lever: increasing the element radiation resistance so that standard feeds require less extreme matching, reporting a two-element end-fire array with spacing $0.103\lambda$ and an impedance bandwidth of about 4.9\% around the 433~MHz band \cite{Best2008}. Together these cases reflect the trade in Fig.~\ref{fig:taxonomy}(c): high directivity can be approached with only two closely spaced resonant elements, but realized gain and usable bandwidth are set by how well mismatch and dissipation are controlled.

The same mechanism becomes explicit in measurements that isolate the loss-limited regime. Altshuler \emph{et al.} study two resonant monopoles at 400~MHz over a finite ground plane while controlling excitation phase and magnitude and sweeping spacing \cite{Altshuler2005}. As spacing is reduced, gain initially increases but eventually drops as the effective radiation resistance becomes small and ohmic loss consumes a large fraction of the accepted power. {This is a useful calibration point for Table~\ref{tab:sota} because it separates ``pattern synthesis'' from ``delivered link benefit'': the directivity trend can persist while $G_{\mathrm{real}}$ collapses once loss and mismatch dominate.}

Parasitic and reactively loaded arrays extend the concept to more elements without a full RF chain per element, but they also make platform coupling difficult to ignore. In the three-element stacked array reported by Haskou \emph{et al.}, the inter-element spacing is about $0.17\lambda$ near 868~MHz and the reported electrical size is $ka\approx 1.6$ \cite{Haskou2017CRPhys}. The reported peak directivity is about 8.5~dBi with radiation efficiency around 35\%, while the measured $|S_{11}|<-10$~dB bandwidth remains in the MHz range, illustrating how quickly impedance bandwidth can collapse in resonant superdirective stacks \cite{Haskou2017CRPhys}. The same work highlights a second practical point: reproducing the fully driven optimal current vector using purely passive parasitic loading can require effective negative resistances, so the feasible design space is constrained by passivity and stability. More broadly, PCB-integration studies show that, for electrically small arrays, the host ground and nearby conductors can reshape the return-current paths that participate in radiation and can strongly alter the achieved directivity \cite{Haskou2016LAWP}. For this reason, Table~\ref{tab:sota} notes when the reported size and performance inherently include a platform or ground plane and should be interpreted as a system result rather than a radiator-only result.

Single-feed mode-engineered radiators illustrate the second route in Fig.~\ref{fig:taxonomy}(b): enforce higher-order radiation content inside one compact body. In the notched dielectric resonator of \cite{KrasnokAPL2014}, the bounding size is subwavelength (on the order of $\sim 0.5\lambda$ across at the operating frequency), and the reported peak simulated directivity reaches $D_{\max}\approx 11$ near 16.4~GHz with high radiation efficiency near the superdirective point \cite{KrasnokAPL2014}. {Related dielectric implementations have also been demonstrated experimentally in compact notch-based resonators, providing a microwave-scale platform that mirrors the multipole-interference picture used in dielectric nanoantenna designs \cite{KrasnokAPL2014}. More recently, superdirectivity of spherical high-index dielectric antennas has been demonstrated experimentally with directivity factors on the order of 10 and high total efficiency at subwavelength size, reinforcing the role of low-loss ceramics and controlled modal interference \cite{Gaponenko2023JAP_SuperdirectiveSpherical}.} This class is important because it separates “number of feeds” from “number of excited modes”: one feed can access a multi-mode mixture when symmetry is intentionally broken and the resonance spectrum is engineered, but the operating band typically remains narrow because the directive response is tied to a small set of resonant modes.

\begin{figure*}[t!]
\centering
\includegraphics[width=1.0\textwidth]{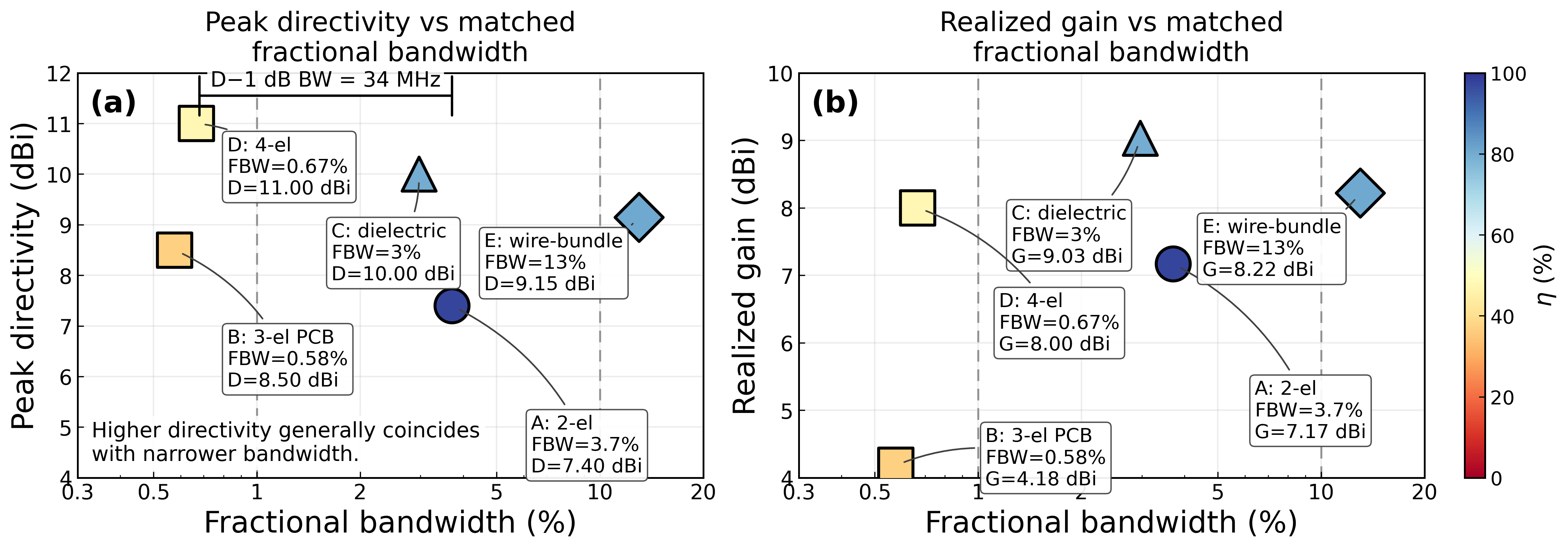}
\caption{\textbf{Directivity--bandwidth--efficiency trade space from reported devices.}
(a) Peak directivity versus {fractional bandwidth (by stated criterion; log scale).}
(b) Realized gain versus {fractional bandwidth (by stated criterion).}
Points correspond to selected devices in Table~\ref{tab:sota}. Bandwidth values are taken as reported and are labeled by criterion when available (e.g., $|S_{11}|<-10$~dB impedance bandwidth or 1~dB gain-drop bandwidth). Panel (b) includes only entries where realized gain and the associated bandwidth criterion are explicitly reported. Marker color indicates reported total or radiation efficiency when available; route classes follow Fig.~\ref{fig:taxonomy}(b). The plot visualizes the trend in Fig.~\ref{fig:taxonomy}(c): higher directivity is often associated with narrower usable bandwidth, while multi-resonant synthesis can partially relax the bandwidth penalty at the cost of increased structural degrees of freedom \cite{YaghjianBest2005,Haskou2017CRPhys,Vovchuk2024}.}
\label{fig:trade}
\end{figure*}

Coupled-resonator dimers provide compact “meta-atom” building blocks in which superdirectivity is set primarily by coupling strength and orientation. Vallecchi \emph{et al.} report a singly driven split-ring-resonator dimer whose overall electrical extent is subwavelength (with individual resonators and separations each on the order of $\sim 0.1\lambda$--$0.2\lambda$), achieving a maximum simulated directivity around $D_{\max}\approx 4.5$ near 2~GHz in the reported configuration \cite{Vallecchi2020}. Related work demonstrates end-fire superdirectivity using dimers of magnetically coupled subwavelength helices \cite{Petrov2020}. These systems again highlight Fig.~\ref{fig:taxonomy}(c): strong coupling can replace complex multiport feeding, but narrowband behavior persists unless multiple resonances are deliberately stacked.

At optical scales, “needle-like” superdirectivity is often closer to a synthesis target than a deployable antenna, but it clarifies what modal control must achieve. Arslanagi\'c and Ziolkowski present a highly subwavelength multilayer cylindrical nanoantenna with overall size on the order of $\lambda_0/10$, optimized to target a prescribed set of scattering coefficients across dipole and higher-order modes \cite{Arslanagic2018PRL}. The lesson is not only that very high directivity requires precise amplitude and phase weighting across several modes, but also that dispersion, absorption, and fabrication tolerances make such weighting difficult to maintain in a robust device. {In many nanophotonic settings the more relevant system metric is coupling into a finite numerical aperture or a guided mode rather than free-space directivity alone; nonetheless, the same multipole-interference mechanism underlies superdirective dielectric nanoantenna concepts \cite{KrasnokNanoscale2014,Kuznetsov2016Science}.}

A defining recent direction is to treat bandwidth and realized gain as explicit design objectives rather than fixed penalties. Vovchuk \emph{et al.} demonstrate a near-field coupled wire-bundle antenna at 6~GHz constrained to a subwavelength volume (approximately a $\sim 0.5\lambda$ bounding cube) and report measured realized gain of 8.22~dBi with about 13\% fractional bandwidth, alongside peak directivity near 9~dBi \cite{Vovchuk2024}. This example is a useful counterpoint to single-resonance supergain peaks: it illustrates how multiple nearby resonant contributions can be positioned to keep realized performance from collapsing away from a single frequency, at the cost of increased structural degrees of freedom and increased sensitivity to fabrication and assembly details.

Other compact microwave implementations sit between classical arrays and mode-engineered radiators. A recent printed example uses a high-permittivity substrate and implements a compact quasi-Yagi-like configuration using capacitively loaded loops together with arc-shaped monopoles to improve end-fire behavior and matching {(reporting a maximum directivity of 7.32~dBi in the published configuration)} \cite{Lu2024LAWP}. {Mixed electric/magnetic moment designs provide another practical route to high realized gain at small size: a recent end-fire Huygens-quadrupole electrically small antenna reports measured $D\approx 7.61$~dBi, $G_{\mathrm{real}}\approx 7.06$~dBi, and overall efficiency $\approx 88\%$ at $ka\approx 0.98$ \cite{Zhang2024HQ}.} {These entries are included in Table~\ref{tab:sota} specifically because they report (or allow extraction of) link-relevant quantities beyond peak $D$.} {Finally, cryogenic system contexts introduce an additional materials lever: when radiation resistance is low and currents are high, reducing dielectric loss and drift can translate directly into higher $\eta_{\mathrm{rad}}$ and more stable realized gain; recent cryogenic dielectric antenna measurements provide device-level data that can be used in such system trade-offs \cite{TorresKrasnok2026Cryogenics}.}

\section{Practical challenges}
The dominant limitations remain bandwidth, loss, and sensitivity, as summarized conceptually in Fig.~\ref{fig:taxonomy}(c) and reflected by the spread of reported points in Fig.~\ref{fig:trade}. Bandwidth is constrained because the current distributions that suppress radiation in most directions store substantial reactive energy and often produce rapidly varying input impedance with frequency. In electrically small radiators, Chu-type scaling provides a baseline trend for why shrinking size increases $Q$ \cite{Chu1948,YaghjianBest2005}, and more general results show that imposing higher directivity typically increases stored energy and further raises $Q$ \cite{YaghjianBest2005}. {In practice, this stored-energy penalty also appears as strong reactive near fields, which makes the achieved pattern and matching sensitive to nearby dielectrics, conductors, and finite ground planes; this is one reason platform co-design is often unavoidable in compact RF systems.} In practice, impedance bandwidth and performance bandwidth should be treated as distinct: a device can satisfy a matching criterion over a narrow band while maintaining its peak realized gain over an even narrower band, particularly when the directive response is tied to a sharp resonance \cite{Yaghjian2008RadioScience,Haskou2017CRPhys}.

Loss and mismatch become critical because superdirective operation often drives radiation resistance downward while keeping currents high. In that regime, small series resistance in conductors, reactive loads, vias, and feed networks can consume a large fraction of accepted power, collapsing realized gain even if directivity remains high. The spacing sweep in \cite{Altshuler2005} makes this mechanism explicit by showing where measured gain departs from the lossless-directivity trend as spacing becomes very small. Similar behavior appears in resonant two-element supergain prototypes, where resonance makes matching feasible but the operating point remains narrowband and sensitive to dissipation \cite{Yaghjian2008RadioScience}. {This loss sensitivity motivates low-loss dielectrics and careful current-path design, and it also motivates exploring temperature as a controllable parameter in certain niches: cryogenic dielectric antenna measurements show that some ceramics exhibit improved loaded $Q$ and small frequency drift at 10~K, while others show strong drift and hysteresis that would undermine stable superdirective operation \cite{TorresKrasnok2026Cryogenics}.} From a system perspective, this is why realized gain and total efficiency are the appropriate performance metrics for comparison and why Fig.~\ref{fig:trade}(b) is often more informative than Fig.~\ref{fig:trade}(a).

Sensitivity is the third recurring practical barrier. Superdirective patterns rely on delicate cancellation among fields radiated by different parts of a compact structure, so small perturbations in geometry, excitation phase/magnitude, nearby objects, and platform currents can cause large pattern changes and rapid degradation of realized gain. This is visible both in tolerance analyses for closely spaced arrays and in measurements on compact prototypes \cite{Altshuler2005,Haskou2017CRPhys}. For electrically small arrays integrated into devices, the ground and host conductors can dominate the final current distribution; PCB-integration studies show that controlling return-current paths on the platform can be as important as the radiator layout itself \cite{Haskou2016LAWP}. As a result, many practical demonstrations favor moderate superdirectivity with higher robustness over extreme peak directivity, consistent with the clustering of measured broadband entries in the mid-range of Fig.~\ref{fig:trade}.

Optical and nanophotonic implementations add frequency-specific constraints. Material dispersion and absorption, nanofabrication tolerances, and environmental loading by substrates and nearby structures complicate the modal alignment needed for stable superdirective behavior. Concept studies such as the multilayer “needle” design clarify the level of modal control required \cite{Arslanagic2018PRL}, but practical figures of merit are often tied to coupling into a guided mode or a finite numerical aperture rather than free-space directivity alone \cite{Kuznetsov2016Science,Novotny2011}. This motivates reporting that connects directionality to the intended system interface, alongside the standard far-field metrics used throughout this review.

\begin{table*}[t!]
\centering
\caption{Representative advanced superdirective antennas and arrays (selected). Reported values follow each source; notes specify whether bandwidth is an impedance or performance bandwidth and whether values are simulated or measured. Size metrics are given in wavelength-scaled form (e.g., $ka$, $d/\lambda$, or a $\lambda$-scaled bounding size).}
\label{tab:sota}
\footnotesize
\setlength{\tabcolsep}{3.2pt}
\renewcommand{\arraystretch}{1.08}
\begin{tabularx}{\textwidth}{@{}L{3.25cm} C{1.25cm} L{2.35cm} L{3.25cm} Y@{}}
\toprule
Platform (structure; route) & Band & Size metric & Key performance & Notes / refs \\
\midrule
2-element resonant end-fire array (driven or parasitic)
& RF/VHF
& $ka\sim 0.7$; $d\sim 0.15\lambda$
& End-fire gain $\sim$6--7~dB (meas); narrowband
& Resonant elements enable measurable supergain with practical feeding; bandwidth and efficiency penalties remain \cite{Yaghjian2008RadioScience}. \\

2-element resonant monopole array (loss-limited study)
& \SI{400}{MHz}
& spacing swept; finite ground plane (system)
& Gain rises as $d/\lambda$ shrinks, then drops when loss dominates (meas)
& Illustrates collapse of realized gain when radiation resistance becomes small and ohmic loss becomes comparable \cite{Altshuler2005}. \\

3-element stacked printed array (loaded/parasitic)
& \SI{868}{MHz}
& $ka\approx 1.6$; $d\sim 0.17\lambda$
& $D\approx$8.8/8.5~dBi (sim/meas); $\eta_{\rm rad}\approx$34.7/37\%;
$|S_{11}|<-10$~dB BW $\approx$1.7/5~MHz (sim/meas)
& Loaded/parasitic synthesis; highlights MHz-scale impedance BW and passivity constraints for ideal parasitic currents \cite{Haskou2017CRPhys}. \\

Notched dielectric resonator (single feed; mode-engineered)
& \SI{16}{GHz}
& bounding size $\sim 0.5\lambda$; $\varepsilon_r\approx 16$
& $D_{\max}\approx 11$ (sim); high $\eta_{\rm rad}$ near operating point; matched BW $\sim$ few \%
& Symmetry breaking enables excitation of higher-order modes with one feed \cite{KrasnokAPL2014}. \\

SRR dimer (coupled resonators; single feed)
& \SI{2}{GHz}
& subwavelength dimer (overall extent $\lesssim 0.4\lambda$)
& $D_{\max}\approx 4.5$ near resonance (sim); narrowband
& Coupling and orientation set effective electric/magnetic moment phasing; validated experimentally \cite{Vallecchi2020}. \\

Helical-element dimer (magnetic coupling; single feed)
& low-GHz
& subwavelength dimer
& End-fire superdirectivity reported; narrowband
& Strong magnetic coupling between resonant helices enables compact end-fire response \cite{Petrov2020}. \\

Multilayer cylindrical ``needle'' nanoantenna (theory)
& optical
& overall size $\sim \lambda_0/10$
& High directivity by design (theory)
& Optimized modal weighting across dipole and higher-order modes; concept study under dispersion/loss constraints \cite{Arslanagic2018PRL}. \\

Multi-resonant wire-bundle (optimized geometry)
& \SI{6}{GHz}
& within $\sim 0.5\lambda$ bounding cube
& $D\approx 9.01/9.15$~dBi and $G_{\rm real}\approx 8.81/8.22$~dBi (sim/meas); FBW $\approx 13$\%
& Broadband realized gain via stacked nearby resonances; optimization-driven synthesis \cite{Vovchuk2024}. \\

Compact microstrip end-fire array (high-$\epsilon_r$; printed)
& microwave
& substrate-miniaturized (high-$\epsilon_r$); ground-backed
& {$D_{\max}\approx 7.32$~dBi (reported); matching emphasized (reported)}
& {Quasi-Yagi-like configuration using capacitively loaded loops and arc monopoles; two-element array on $\epsilon_r=24$ substrate \cite{Lu2024LAWP}.} \\

 {Huygens-quadrupole end-fire ESA (single feed; mixed $p$/$m$ moments)}
& {microwave}
& {$ka\approx 0.98$}
& {$D\approx 7.61$~dBi, $G_{\rm real}\approx 7.06$~dBi, OE $\approx 88.1\%$ (meas)}
& {Two near-field resonant parasitic Huygens dipole radiators arranged in an end-fire quadrupole; single-feed prototype validated experimentally \cite{Zhang2024HQ}.} \\
\bottomrule
\end{tabularx}
\end{table*}

%%%%%%%%%%

\section{Outlook}
The trade space in Fig.~\ref{fig:taxonomy}(c) and the reported points in Fig.~\ref{fig:trade} indicate where progress is most valuable: improving \emph{realized} gain over a stated bandwidth at a stated electrical size, while keeping performance stable under realistic perturbations. When superdirectivity is pushed aggressively, the limiting factor is often not the ability to synthesize a sharp pattern at a single frequency, but the combination of mismatch, dissipation, and sensitivity that reduces delivered link benefit. {A useful near-term target is therefore not ``more peak directivity,'' but higher realized gain subject to explicit constraints on bandwidth, efficiency, and robustness, reported in a way that allows fair comparison across platforms and that makes the system-level cost of any tuning or control mechanism explicit.}

A first open problem is measurement and reporting consistency for electrically small superdirective systems. For many designs the ``antenna'' is inseparable from a ground plane, a housing, or a host PCB that carries return currents and re-radiates. In that regime, radiator-only size and system size can differ materially, and both can be relevant depending on the application. A practical standard for this review is to report $ka$ (or a $\lambda$-scaled bounding size) for the radiating system that was actually tested, and to state clearly whether the platform is included. The same clarity is needed for bandwidth: impedance bandwidth and performance bandwidth frequently differ for narrow resonant peaks, so a single reported bandwidth number is ambiguous unless the criterion is specified. {When bandwidth is reported in absolute units, the reported center frequency should also be stated so that fractional bandwidth can be reconstructed unambiguously, and plots should avoid labels such as ``matched bandwidth'' unless the definition is strictly an impedance-matching criterion at a specified reference plane.} Finally, robustness needs a reportable definition. For superdirective devices, a natural metric is a worst-case or percentile realized gain across the target band under small but stated perturbations in geometry, excitation, and environment, because that directly captures the cancellation sensitivity that limits deployable performance.

A second open problem is controllable tuning that preserves efficiency. Many compact arrays and mode-engineered radiators exhibit a strong directive peak over a narrow band; in practice, tuning is often more valuable than attempting to make that peak intrinsically wide. Electronically steerable parasitic array radiator (ESPAR) antennas illustrate a mature control mechanism in which one driven element is surrounded by parasitic elements with variable reactive loads, enabling beam steering with a single RF chain \cite{OhiraGyoda2000ESPAR,OhiraIigusa2004ESPAR}. The same idea applies to superdirective arrays and dimers, but the requirements are stricter: the tuning network becomes part of the radiator, so its loss, nonlinearity, bias dependence, and stability directly affect realized gain and pattern integrity. {In this regime, it is helpful to treat tuning as part of the ``antenna'' definition for reporting: the relevant efficiency is the end-to-end total efficiency that includes the dissipative loss of bias networks, tuning elements, and any packaging needed to keep the operating point stable.} A central engineering challenge is to implement low-loss, low-noise, and drift-tolerant tuning while keeping the operating point inside the narrow ``useful'' region of Fig.~\ref{fig:trade}.

 {A complementary lever is to reduce dissipative loss rather than only reshaping currents. Superdirectivity often pushes current amplitudes up while radiation resistance drops, so conductor and dielectric losses become first-order determinants of $G_{\mathrm{real}}$. This makes materials and temperature control unusually impactful compared with more conventional apertures. In specialized systems, cryogenic operation can reduce loss and improve stability of certain dielectric resonators and feed networks, which can help close the gap between peak directivity and realized gain without changing the radiating geometry. For example, cryogenic dielectric resonator antennas based on ZST ceramics have been shown to exhibit small frequency drift and improved loaded $Q$ at 10~K, enabling milliwatt-level through-window wireless links across a thermal boundary \cite{TorresKrasnok2026Cryogenics}. This line of work also highlights a practical caution: not all high-$Q$ ceramics behave benignly at low temperature, and hysteresis or strong drift can defeat the goal of stable superdirective operation even if room-temperature loss tangents look favorable \cite{TorresKrasnok2026Cryogenics}. For cryogenic quantum hardware, where wiring heat load and channel count are central constraints, antenna-based wireless links inside a cryostat are emerging as an active systems topic, and they provide a concrete use-case where narrowband but highly efficient antennas may be preferable to broadband solutions \cite{Centritto2025CryoCMOS}.}

 {A third direction is to use time modulation to relax bandwidth limits that apply to linear time-invariant (LTI) passive matching. Since the Chu and Bode--Fano bounds are derived under LTI assumptions, time-varying (linear time-varying, LTV) matching networks and loads offer a principled route to improved bandwidth at fixed electrical size, at the cost of additional control complexity and, in general, frequency conversion. In Floquet impedance matching, a temporally modulated matching network is tailored so that the effective matching condition is satisfied over a broader band than would be possible in the LTI passive setting \cite{LiMekawyAlu2019PRL}. Related parametric approaches modulate reactive loads to reshape both the effective input response and the radiative interaction of an electrically small radiator, enabling enhanced radiation compared with stationary loading \cite{MekawyLiRadiAlu2021PRAppl}. Time-modulated radiators have also been used to access functionalities not available in reciprocal LTI antennas, such as asymmetric transmit/receive behavior \cite{HadadSoricAlu2016PNAS}. For superdirective antennas, the key opportunity is that time modulation can supply additional degrees of freedom to keep a desired current or modal superposition ``on target'' across frequency, rather than only at one resonance. The key reporting challenge is that conventional $G_{\mathrm{real}}$ does not capture where power goes spectrally: time modulation can redistribute power into sidebands, and it may require non-negligible modulation power. A practical reporting standard for time-modulated superdirective antennas is therefore to state (i) carrier-only realized gain and bandwidth, (ii) total radiated power partition across the generated tones, and (iii) a system efficiency that includes the energy delivered by the modulation mechanism, alongside the conventional RF-port definitions \cite{HayranMonticone2023APM}.}

A fourth direction is to close the gap between directivity and realized gain by reducing loss and simplifying matching. In several entries in Table~\ref{tab:sota}, the measured pattern sharpening does not translate into comparable link improvement because accepted power is reflected by mismatch or dissipated in conductors, vias, lumped loads, and matching networks. This motivates designs with self-resonant elements that present higher radiation resistance, matching that is integrated into the radiator rather than added externally, and current distributions that avoid extreme amplitudes. Active matching with non-Foster elements can extend impedance matching beyond passive limits \cite{SussmanFortRudish2009,AlbarracinVargas2016DesignNonFoster}, but the antenna and active network must be treated as a single system because stability and placement sensitivity can dominate performance \cite{AlbarracinVargas2014SensitivityNonFoster,CareyAberle2025PostMortemNonFoster}. {Time-modulated matching and cryogenic loss reduction should be viewed in the same system-level frame as non-Foster matching: the question is not only whether the impedance curve looks flatter, but whether the delivered link metric improves after accounting for added noise, added power consumption, bias drift, and spectral constraints.} In many applications, the most reliable system outcome may come from moderate superdirectivity with high total efficiency, achieved with passive matching that is robust to tolerances and platform coupling.

Bandwidth extension through mode stacking and geometry synthesis is likely to expand as inverse design tools mature. The multi-resonant wire-bundle example in Table~\ref{tab:sota} shows that several nearby resonant contributions can be placed close enough to maintain high realized gain over a wider band \cite{Vovchuk2024}. As design spaces become higher-dimensional, surrogate modeling and data-driven optimization can reduce the number of full-wave evaluations \cite{Inman2004PIER,Gajbhiye2025DiscoverElectronics}. The practical risk is fragility: optimizers can discover narrow optima that collapse under fabrication spread, bias drift, or modest platform changes. This makes tolerance-aware optimization and yield-driven objectives central for future work, because they bias the search toward designs whose realized gain remains high across the required band for realistic perturbation distributions \cite{KozielPietrenko2022ToleranceOptimization}. {An important next step is co-design across physics layers: geometry, matching/tuning, and (when used) time modulation should be optimized together under explicit constraints on loss, spectral purity, and stability, rather than validated sequentially after the fact.}

Frequency regime matters, but the underlying design logic stays the same. In RF and VHF, severe size constraints coincide with finite ground planes, high current levels, and unforgiving matching networks, so conservative superdirectivity combined with tuning and careful platform co-design often provides the best system value. In microwave and mm-wave regimes, manufacturing tolerances are smaller relative to wavelength and integration is easier, so multi-resonant and optimization-driven designs become more viable, provided that robustness is designed in rather than checked afterward. {In cryogenic platforms, the balance can shift: extremely low available cooling power and wiring congestion can make wireless links attractive, and the combination of low-loss dielectrics, stable resonances, and modest superdirectivity may be more valuable than pursuing wide instantaneous bandwidth \cite{TorresKrasnok2026Cryogenics,Centritto2025CryoCMOS}.} In optics, low-loss dielectrics and mode interference remain promising, but the most relevant figure of merit is often coupling into a guided mode or collection numerical aperture rather than free-space directivity alone, so superdirective behavior is best evaluated as part of a photonic system interface.

The bounds and trade-offs summarized in Fig.~\ref{fig:taxonomy}(c) remain the right physical framing. {What is changing is the set of practical levers that can move devices toward higher realized gain at a stated bandwidth and size: low-loss materials (including cryogenic operation in niche systems), tuning networks engineered as part of the radiator, and time-varying (Floquet/parametric) matching and loading that can relax LTI bandwidth constraints when their spectral and power costs are reported transparently \cite{LiMekawyAlu2019PRL,MekawyLiRadiAlu2021PRAppl,HayranMonticone2023APM}.}

\section{Conclusion}
Superdirective antennas remain constrained by the same mechanisms that historically made them difficult to deploy: large stored energy, narrow usable bandwidth, and strong sensitivity to perturbations. The key practical lesson reinforced by Table~\ref{tab:sota} and Fig.~\ref{fig:trade} is that peak directivity is rarely predictive of link benefit by itself. Repeatable progress is clearest when results are reported and compared using realized gain (or total efficiency times directivity), a stated bandwidth criterion, and an explicit size definition that matches the radiating system that was actually tested.

Across the main implementation routes, the physics is consistent even when the hardware differs. Resonant two-element arrays remain the cleanest benchmark because they expose mismatch and loss penalties directly and make sensitivity to excitation errors and spacing explicit. Parasitic loading and tunable reactances reduce feed count, but they increase reliance on coupling, platform currents, and the loss and stability of the tuning network. Mode-engineered radiators demonstrate that a single feed can access a controlled multipole mixture, but the operating band remains tied to the stability of a small set of resonant modes. Multi-resonant and optimization-driven geometries show that usable bandwidth can be extended by adding degrees of freedom, provided robustness is treated as a design objective rather than a post-check.

Two emerging levers broaden the perspective beyond purely geometric synthesis. First, loss reduction and stability can be pursued through materials and temperature control in niche systems where narrowband operation is acceptable; cryogenic dielectric antenna measurements show that certain ceramics can improve loaded $Q$ and maintain small drift at 10~K, enabling low-power wireless links across a thermal boundary and suggesting a route to higher realized gain when radiation resistance is low and currents are high \cite{TorresKrasnok2026Cryogenics}. Second, time-varying loading and matching can relax bandwidth limits derived under linear time-invariant assumptions; Floquet impedance matching and related parametric schemes have demonstrated pathways to exceed classical small-antenna limits, but they also introduce frequency conversion and additional system costs that must be reported explicitly (e.g., carrier-only gain, sideband power partition, and modulation power) \cite{LiMekawyAlu2019PRL,MekawyLiRadiAlu2021PRAppl,HayranMonticone2023APM}. For superdirective antennas, these tools are most valuable when they are used to keep a desired current or modal superposition stable across frequency and loading, rather than to maximize a single-frequency directivity peak.

The practical route forward is therefore consistent and measurable: maximize realized gain over the required band at the required electrical size; report size, reference plane, and bandwidth definitions unambiguously (including when the platform is part of the radiating system); and quantify sensitivity with a stated perturbation model. Progress should be judged by moving devices into regions of Fig.~\ref{fig:trade} where realized gain remains high under fabrication spread, bias drift, and host-environment loading, and by adopting reporting practices that make fair comparison possible across LTI, actively matched, and time-modulated implementations.

\section*{Acknowledgment}
The author acknowledges financial support from the U.S. Department of Energy (DoE) and the U.S. Air Force Office of Scientific Research (AFOSR).

\bibliographystyle{IEEEtran}
\bibliography{references}

@book{Balanis2016,
  author    = {Constantine A. Balanis},
  title     = {Antenna Theory: Analysis and Design},
  edition   = {4},
  publisher = {Wiley},
  year      = {2016}
}

@article{HansenWoodyard1938,
  author  = {Hansen, W. W. and Woodyard, J. R.},
  title   = {A New Principle in Directional Antenna Design},
  journal = {Proceedings of the IRE},
  volume  = {26},
  number  = {3},
  pages   = {333--345},
  year    = {1938},
  doi     = {10.1109/JRPROC.1938.228128}
}

@article{Uzkov1946,
  author  = {A. I. Uzkov},
  title   = {An Approach to the Problem of Optimum Directive Antenna Design},
  journal = {Doklady Akademii Nauk SSSR},
  volume  = {53},
  number  = {1},
  pages   = {35--38},
  year    = {1946},
  note    = {in Russian}
}

@article{Wheeler1947,
  author  = {H. A. Wheeler},
  title   = {Fundamental Limitations of Small Antennas},
  journal = {Proceedings of the IRE},
  volume  = {35},
  number  = {12},
  pages   = {1479--1484},
  year    = {1947}
}

@article{Chu1948,
  author  = {L. J. Chu},
  title   = {Physical Limitations of Omni-Directional Antennas},
  journal = {Journal of Applied Physics},
  volume  = {19},
  number  = {12},
  pages   = {1163--1175},
  year    = {1948},
  doi     = {10.1063/1.1715038}
}

@article{Harrington1958,
  author  = {R. F. Harrington},
  title   = {On the Gain and Beamwidth of Directional Antennas},
  journal = {IRE Transactions on Antennas and Propagation},
  volume  = {6},
  number  = {3},
  pages   = {219--225},
  year    = {1958}
}

@article{Harrington1960,
  author  = {R. F. Harrington},
  title   = {Effect of Antenna Size on Gain, Bandwidth, and Efficiency},
  journal = {Journal of Research of the National Bureau of Standards, Section D: Radio Propagation},
  volume  = {64D},
  number  = {1},
  pages   = {1--12},
  year    = {1960}
}

@article{Fano1950,
  author  = {R. M. Fano},
  title   = {Theoretical Limitations on the Broadband Matching of Arbitrary Impedances},
  journal = {Journal of the Franklin Institute},
  volume  = {249},
  number  = {1},
  pages   = {57--83},
  year    = {1950},
  doi     = {10.1016/0016-0032(50)90006-8}
}

@article{YaghjianBest2005,
  author  = {A. D. Yaghjian and S. R. Best},
  title   = {Impedance, Bandwidth, and {Q} of Antennas},
  journal = {IEEE Transactions on Antennas and Propagation},
  volume  = {53},
  number  = {4},
  pages   = {1298--1324},
  year    = {2005},
  doi     = {10.1109/TAP.2005.844443}
}

@article{Altshuler2005,
  author  = {E. E. Altshuler and T. H. O'Donnell and A. D. Yaghjian and S. R. Best},
  title   = {A Monopole Superdirective Array},
  journal = {IEEE Transactions on Antennas and Propagation},
  volume  = {53},
  number  = {8},
  pages   = {2653--2661},
  year    = {2005},
  doi     = {10.1109/TAP.2005.851810}
}

@article{Best2008,
  author  = {S. R. Best and E. E. Altshuler and A. D. Yaghjian and J. M. McGinthy and T. H. O'Donnell},
  title   = {An Impedance-Matched 2-Element Superdirective Array},
  journal = {IEEE Antennas and Wireless Propagation Letters},
  volume  = {7},
  pages   = {302--305},
  year    = {2008},
  doi     = {10.1109/LAWP.2008.921372}
}

@article{Yaghjian2008RadioScience,
  author  = {A. D. Yaghjian and T. H. O'Donnell and E. E. Altshuler and S. R. Best},
  title   = {Electrically Small Supergain End-Fire Arrays},
  journal = {Radio Science},
  volume  = {43},
  pages   = {RS3002},
  year    = {2008},
  doi     = {10.1029/2007RS003747}
}

@article{Haskou2015,
  author  = {A. Haskou and A. Sharaiha and S. Collardey},
  title   = {Design of Small Parasitic Loaded Superdirective End-Fire Antenna Arrays},
  journal = {IEEE Transactions on Antennas and Propagation},
  volume  = {63},
  number  = {12},
  pages   = {5456--5464},
  year    = {2015},
  doi     = {10.1109/TAP.2015.2496112}
}

@article{Haskou2017CRPhys,
  author  = {A. Haskou and A. Sharaiha and S. Collardey},
  title   = {Theoretical and Practical Limits of Superdirective Antenna Arrays},
  journal = {Comptes Rendus Physique},
  volume  = {18},
  pages   = {118--124},
  year    = {2017},
  doi     = {10.1016/j.crhy.2016.11.003}
}

@article{Lu2024LAWP,
  author  = {P. Lu and Z. Liu and E. Lei and Z. Chen and X. Zheng and C. Song and G. A. E. Vandenbosch},
  title   = {Compact Superdirective Microstrip Antenna Array Using Capacitively Loaded Loops on High Dielectric Substrate},
  journal = {IEEE Antennas and Wireless Propagation Letters},
  volume  = {23},
  number  = {4},
  pages   = {1351--1355},
  year    = {2024},
  doi     = {10.1109/LAWP.2024.3355703}
}

@article{KrasnokAPL2014,
  author  = {A. E. Krasnok and D. S. Filonov and C. R. Simovski and Y. S. Kivshar and P. A. Belov},
  title   = {Experimental Demonstration of Superdirective Dielectric Antenna},
  journal = {Applied Physics Letters},
  volume  = {104},
  number  = {13},
  pages   = {133502},
  year    = {2014},
  doi     = {10.1063/1.4869817}
}

@article{KrasnokNanoscale2014,
  author  = {A. E. Krasnok and C. R. Simovski and P. A. Belov and Y. S. Kivshar},
  title   = {Superdirective Dielectric Nanoantennas},
  journal = {Nanoscale},
  volume  = {6},
  number  = {13},
  pages   = {7354--7361},
  year    = {2014},
  doi     = {10.1039/C4NR01231C}
}

@article{Vallecchi2020,
  author  = {A. Vallecchi and A. Radkovskaya and L. Li and G. Faulkner and C. J. Stevens and E. Shamonina},
  title   = {Superdirective Dimers of Coupled Self-Resonant Split Ring Resonators: Analytical Modelling and Numerical and Experimental Validation},
  journal = {Scientific Reports},
  volume  = {10},
  number  = {1},
  pages   = {274},
  year    = {2020},
  doi     = {10.1038/s41598-019-56988-6}
}

@article{Petrov2020,
  author  = {P. Petrov and A. P. Hibbins and J. R. Sambles},
  title   = {Microwave Superdirectivity with Dimers of Helical Elements},
  journal = {Physical Review Applied},
  volume  = {13},
  number  = {4},
  pages   = {044012},
  year    = {2020},
  doi     = {10.1103/PhysRevApplied.13.044012}
}

@article{Vovchuk2024,
  author  = {D. Vovchuk and G. Uziel and A. Machnev and J. Porins and V. Bobrovs and P. Ginzburg},
  title   = {Genetically Synthesized Supergain Broadband Wire-Bundle Antenna},
  journal = {Communications Engineering},
  volume  = {3},
  number  = {1},
  pages   = {101},
  year    = {2024},
  doi     = {10.1038/s44172-024-00235-y}
}

@article{Arslanagic2018PRL,
  author  = {S. Arslanagi{\'c} and R. W. Ziolkowski},
  title   = {Highly Subwavelength, Superdirective Cylindrical Nanoantenna},
  journal = {Physical Review Letters},
  volume  = {120},
  number  = {23},
  pages   = {237401},
  year    = {2018},
  doi     = {10.1103/PhysRevLett.120.237401}
}

@article{Kuznetsov2016Science,
  author  = {A. I. Kuznetsov and A. E. Miroshnichenko and M. L. Brongersma and Y. S. Kivshar and B. Lukyanchuk},
  title   = {Optically Resonant Dielectric Nanostructures},
  journal = {Science},
  volume  = {354},
  number  = {6314},
  pages   = {aag2472},
  year    = {2016},
  doi     = {10.1126/science.aag2472}
}

@article{Novotny2011,
  author  = {L. Novotny and N. van Hulst},
  title   = {Antennas for Light},
  journal = {Nature Photonics},
  volume  = {5},
  number  = {2},
  pages   = {83--90},
  year    = {2011},
  doi     = {10.1038/nphoton.2010.237}
}

@article{Biagioni2012,
  author  = {P. Biagioni and J.-S. Huang and B. Hecht},
  title   = {Nanoantennas for Visible and Infrared Radiation},
  journal = {Reports on Progress in Physics},
  volume  = {75},
  number  = {2},
  pages   = {024402},
  year    = {2012},
  doi     = {10.1088/0034-4885/75/2/024402}
}

@article{Zhang2024HQ,
  author  = {Z. Zhang and M. Li and Q. Dai and R. W. Ziolkowski},
  title   = {Superdirective, Electrically Small, Endfire-Radiating Huygens Quadrupole Antenna},
  journal = {IEEE Transactions on Antennas and Propagation},
  volume  = {72},
  number  = {10},
  pages   = {7615--7627},
  year    = {2024},
  doi     = {10.1109/TAP.2024.3451155}
}

@standard{IEEE1452013,
  title        = {{IEEE Std 145-2013}: IEEE Standard for Definitions of Terms for Antennas},
  organization = {IEEE},
  year         = {2014},
  note         = {Approved 11 Dec. 2013; published 11 Mar. 2014}
}

@article{GustafssonNordebo2013,
  author  = {Gustafsson, Mats and Nordebo, Sven},
  title   = {Optimal Antenna Currents for {Q}, Superdirectivity, and Radiation Patterns Using Convex Optimization},
  journal = {IEEE Transactions on Antennas and Propagation},
  volume  = {61},
  number  = {3},
  pages   = {1109--1118},
  year    = {2013},
  doi     = {10.1109/TAP.2012.2227656}
}

@article{GustafssonCapek2019,
  author  = {Gustafsson, Mats and Capek, Miloslav},
  title   = {Maximum Gain, Effective Area, and Directivity},
  journal = {IEEE Transactions on Antennas and Propagation},
  volume  = {67},
  number  = {8},
  pages   = {5282--5293},
  year    = {2019},
  doi     = {10.1109/TAP.2019.2916760}
}

@article{Jonsson2017QDirectivity,
  author  = {Jonsson, B. L. G. and Shi, Shuai and Wang, Lei and Ferrero, Fabien and Lizzi, Leonardo},
  title   = {On Methods to Determine Bounds on the {Q}-Factor for a Given Directivity},
  journal = {IEEE Transactions on Antennas and Propagation},
  volume  = {65},
  number  = {11},
  pages   = {5686--5696},
  year    = {2017},
  doi     = {10.1109/TAP.2017.2748383}
}

@article{Haskou2016LAWP,
  author  = {Haskou, Abdullah and Sharaiha, Ala and Collardey, Sylvain},
  title   = {Integrating Superdirective Electrically Small Antenna Arrays in PCBs},
  journal = {IEEE Antennas and Wireless Propagation Letters},
  volume  = {15},
  pages   = {24--27},
  year    = {2016},
  doi     = {10.1109/LAWP.2015.2425913}
}

@inproceedings{Jaafar2018URSI,
  author    = {Jaafar, Hussein and Collardey, Sylvain and Sharaiha, Ala},
  title     = {Network Characteristic Modes Optimisation for Wideband and Superdirective Small Antennas},
  booktitle = {2018 2nd URSI Atlantic Radio Science Meeting (AT-RASC)},
  year      = {2018},
  doi       = {10.23919/URSI-AT-RASC.2018.8471637}
}

@article{HarringtonMautz1971,
  author  = {Harrington, R. F. and Mautz, J. R.},
  title   = {Computation of Characteristic Modes for Conducting Bodies},
  journal = {IEEE Transactions on Antennas and Propagation},
  volume  = {19},
  number  = {5},
  pages   = {629--639},
  year    = {1971},
  doi     = {10.1109/TAP.1971.1139990}
}

@article{GarbaczPozar1982,
  author  = {Garbacz, R. J. and Pozar, D. M.},
  title   = {Antenna Shape Synthesis Using Characteristic Modes},
  journal = {IEEE Transactions on Antennas and Propagation},
  volume  = {30},
  number  = {3},
  pages   = {340--350},
  year    = {1982},
  doi     = {10.1109/TAP.1982.1142820}
}

@article{Shamonina2002JAP,
  author  = {Shamonina, E. and Kalinin, V. A. and Ringhofer, K. H. and Solymar, L.},
  title   = {Magnetoinductive waves in one, two, and three dimensions},
  journal = {Journal of Applied Physics},
  volume  = {92},
  number  = {10},
  pages   = {6252--6261},
  year    = {2002},
  doi     = {10.1063/1.1510945}
}

@article{SussmanFortRudish2009,
  author  = {Sussman-Fort, Stephen E. and Rudish, Ronald M.},
  title   = {Non-Foster Impedance Matching of Electrically-Small Antennas},
  journal = {IEEE Transactions on Antennas and Propagation},
  volume  = {57},
  number  = {8},
  pages   = {2230--2241},
  year    = {2009},
  month   = aug,
  doi     = {10.1109/TAP.2009.2024494}
}

@article{AlbarracinVargas2014SensitivityNonFoster,
  author  = {Albarrac{\'i}n-Vargas, Fernando and Ugarte-Mu{\~n}oz, Eduardo and Gonz{\'a}lez-Posadas, Vicente and Segovia-Vargas, Daniel},
  title   = {Sensitivity Analysis for Active Matched Antennas With Non-Foster Elements},
  journal = {IEEE Transactions on Antennas and Propagation},
  volume  = {62},
  number  = {12},
  pages   = {6040--6048},
  year    = {2014},
  month   = dec,
  doi     = {10.1109/TAP.2014.2364811}
}

@article{AlbarracinVargas2016DesignNonFoster,
  author  = {Albarrac{\'i}n-Vargas, Fernando and Gonz{\'a}lez-Posadas, Vicente and Herraiz-Mart{\'i}nez, Francisco Javier and Segovia-Vargas, Daniel},
  title   = {Design Method for Actively Matched Antennas With Non-Foster Elements},
  journal = {IEEE Transactions on Antennas and Propagation},
  volume  = {64},
  number  = {9},
  pages   = {4118--4123},
  year    = {2016},
  month   = sep,
  doi     = {10.1109/TAP.2016.2583482}
}

@misc{CareyAberle2025PostMortemNonFoster,
  author       = {Carey, Robert and Aberle, James},
  title        = {A Post-Mortem on Non-Foster Matching Networks For Electrically Small Antennas},
  howpublished = {TechRxiv preprint},
  year         = {2025},
  month        = feb,
  doi          = {10.36227/techrxiv.173950943.34303942.v1},
  note         = {Preprint; posted 14 Feb 2025}
}

@conference{OhiraGyoda2000ESPAR,
  author    = {Ohira, Takashi and Gyoda, Koichi},
  title     = {Electronically Steerable Passive Array Radiator Antennas for Low-Cost Analog Adaptive Beamforming},
  booktitle = {Proc. IEEE International Conference on Phased Array Systems and Technology},
  address   = {Dana Point, CA, USA},
  pages     = {101--104},
  year      = {2000}
}

@article{OhiraIigusa2004ESPAR,
  author  = {Ohira, Takashi and Iigusa, Kyouichi},
  title   = {Electronically Steerable Parasitic Array Radiator Antenna},
  journal = {Electronics and Communications in Japan (Part II: Electronics)},
  volume  = {87},
  number  = {10},
  pages   = {25--45},
  year    = {2004},
  month   = oct,
  doi     = {10.1002/ecjb.20081}
}

@article{KozielPietrenko2022ToleranceOptimization,
  author  = {Koziel, Slawomir and Pietrenko-D{\k a}browska, Anna},
  title   = {Tolerance Optimization of Antenna Structures by Means of Response Feature Surrogates},
  journal = {IEEE Transactions on Antennas and Propagation},
  volume  = {70},
  number  = {11},
  pages   = {10988--10997},
  year    = {2022},
  month   = nov,
  doi     = {10.1109/TAP.2022.3187665}
}

@article{Inman2004PIER,
  author  = {Inman, M. J. and Earwood, J. M. and Elsherbeni, A. Z. and Smith, C. E.},
  title   = {Bayesian Optimization Techniques for Antenna Design},
  journal = {Progress In Electromagnetics Research},
  volume  = {49},
  pages   = {71--86},
  year    = {2004},
  doi     = {10.2528/PIER04021302}
}

@article{Gajbhiye2025DiscoverElectronics,
  author  = {Gajbhiye, Pradnya A. and Singh, Satya P. and Sharma, Madan Kumar},
  title   = {A comprehensive review of {AI} and machine learning techniques in antenna design optimization and measurement},
  journal = {Discover Electronics},
  year    = {2025},
  doi     = {10.1007/s44291-025-00084-9}
}

@article{TorresKrasnok2026Cryogenics,
  title   = {A cryogenic dielectric antenna for wireless sensing and interfacing outside the 10 K environment},
  author  = {Torres, Ingrid and Krasnok, Alex},
  journal = {Cryogenics},
  volume  = {155},
  pages   = {104286},
  year    = {2026},
  month   = mar,
  doi     = {10.1016/j.cryogenics.2026.104286},
  url     = {https://doi.org/10.1016/j.cryogenics.2026.104286}
}

@article{Gaponenko2023JAP_SuperdirectiveSpherical,
  author  = {Gaponenko, Roman and Sidorenko, Mikhail S. and Zhirihin, Dmitry and Rasskazov, Ilia L. and Moroz, Alexander and Ladutenko, Konstantin and Belov, Pavel and Shcherbakov, Alexey},
  title   = {Experimental demonstration of superdirective spherical dielectric antenna},
  journal = {Journal of Applied Physics},
  volume  = {134},
  number  = {1},
  pages   = {014901},
  year    = {2023},
  doi     = {10.1063/5.0155677},
}

@article{LiMekawyAlu2019PRL,
  author  = {Li, Huanan and Mekawy, Ahmed and Al{\`u}, Andrea},
  title   = {Beyond {C}hu's Limit with {F}loquet Impedance Matching},
  journal = {Physical Review Letters},
  volume  = {123},
  number  = {16},
  pages   = {164102},
  year    = {2019},
  month   = oct,
  doi     = {10.1103/PhysRevLett.123.164102},
}

@article{MekawyLiRadiAlu2021PRAppl,
  author  = {Mekawy, Ahmed and Li, Huanan and Ra'di, Younes and Al{\`u}, Andrea},
  title   = {Parametric Enhancement of Radiation from Electrically Small Antennas},
  journal = {Physical Review Applied},
  volume  = {15},
  number  = {5},
  pages   = {054063},
  year    = {2021},
  month   = may,
  doi     = {10.1103/PhysRevApplied.15.054063},
}

@article{HadadSoricAlu2016PNAS,
  author  = {Hadad, Yakir and Soric, Jason C. and Al{\`u}, Andrea},
  title   = {Breaking temporal symmetries for emission and absorption},
  journal = {Proceedings of the National Academy of Sciences of the United States of America},
  volume  = {113},
  number  = {13},
  pages   = {3471--3475},
  year    = {2016},
  month   = mar,
  doi     = {10.1073/pnas.1517363113},
}

@article{HayranMonticone2023APM,
  author  = {Hayran, Zeki and Monticone, Francesco},
  title   = {Using Time-Varying Systems to Challenge Fundamental Limitations in Electromagnetics: {O}verview and summary of applications},
  journal = {IEEE Antennas and Propagation Magazine},
  volume  = {65},
  number  = {4},
  pages   = {29--38},
  year    = {2023},
  doi     = {10.1109/MAP.2023.3236275},
}

@article{Centritto2025CryoCMOS,
  author  = {Centritto, Viviana and Bandara, Ama and Deng, Heqi and Babaie, Masoud and Vinogradov, Evgenii and Abadal, Sergi and Alarcon, Eduard},
  title   = {Cryo-CMOS Antenna for Wireless Communications within a Quantum Computer Cryostat},
  journal = {arXiv},
  year    = {2025},
  note    = {arXiv:2510.13627},
  doi     = {10.48550/arXiv.2510.13627},
}

@article{Abulgasem2021CubesatAntennaReview,
  author  = {Abulgasem, Suhila and Tubbal, Faisel E. M. and Raad, Raad and Theoharis, Panagiotis Ioannis and Liu, Sining and Iranmanesh, Saeid},
  title   = {Antenna Designs for CubeSats: {A} Review},
  journal = {IEEE Access},
  volume  = {9},
  pages   = {45289--45324},
  year    = {2021},
  doi     = {10.1109/ACCESS.2021.3066632},
  url     = {https://doi.org/10.1109/ACCESS.2021.3066632}
}

@article{Dai2013DirectionalAntennas,
  author  = {Dai, Hongning and Ng, Kam-Wing and Li, Minglu and Wu, Min-You},
  title   = {An overview of using directional antennas in wireless networks},
  journal = {International Journal of Communication Systems},
  volume  = {26},
  number  = {4},
  pages   = {413--448},
  year    = {2013},
  doi     = {10.1002/dac.1348},
  url     = {https://doi.org/10.1002/dac.1348}
}

\begin{IEEEbiography}[{\includegraphics[width=1in,height=1.25in,clip,keepaspectratio]{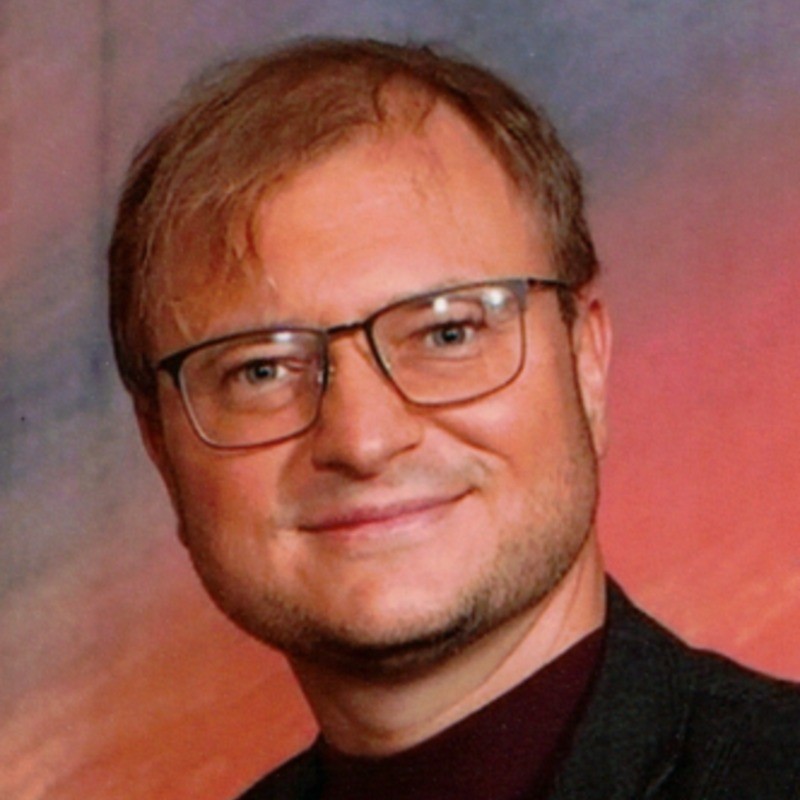}}]{Alex Krasnok}
Alex Krasnok is an Assistant Professor with the Department of Electrical and Computer Engineering, Florida International University (FIU), Miami, FL, USA, with a courtesy appointment in the Knight Foundation School of Computing and Information Sciences. He directs the Quantum Technology and Metamaterials (QTM) Laboratory at FIU. His research spans electrodynamics and wave physics with emphasis on electrically small and resonant antennas, metasurfaces and metadevices, and quantum and optical sensing.

Dr. Krasnok received the M.S. degree in quantum optics from Far Eastern Federal University, Russia, and the Ph.D. degree in photonics and quantum optics from ITMO University, Russia. He was a Postdoctoral Fellow in photonics with The University of Texas at Austin. He has authored more than 200 peer-reviewed publications and has led federally funded research projects supported by the U.S. Department of Energy, the U.S. Air Force Office of Scientific Research, the National Science Foundation, and the Office of Naval Research. His recognitions include the 2024 EurAAP Leopold B.\ Felsen Award for excellence in electrodynamics and FIU Top Scholar recognition for the 2024--2025 academic year.

\end{IEEEbiography}

\end{document}